# Squeezed States Generation using Cryogenic InP HEMT Transistor Nonlinearity

Ahmad Salmanogli

Cankaya University, Engineering faculty, Electrical and Electronic Department, Ankara, Turkey

**Abstract:** This study focuses on generating and manipulating squeezed states with two external oscillators coupled by an InP HEMT transistor operating at cryogenic temperatures. First, the small-signal nonlinear model of the transistor at high frequency at 5 K is analyzed using quantum theory, and the related Lagrangian is theoretically derived. Subsequently, the total quantum Hamiltonian of the system is derived using Legendre transformation. The Hamiltonian of the system includes linear and nonlinear terms, by which the effects on the time evolution of the states are studied. The main result shows that the squeezed state can be generated owing to the nonlinearity of the transistor, and more importantly, it can be manipulated by some specific terms introduced in the nonlinear Hamiltonian. In fact, the nonlinearity of the transistors induces some effects such as capacitance, inductance, and second-order transconductance, by which the properties of the external oscillators are changed. These changes may lead to squeezing or manipulation of the parameters related to squeezing in the oscillators. In addition, it is theoretically derived that the circuit can generate two-mode squeezing. Finally, second-order correlation (photon counting statistics) is studied as a complementary task, and the results demonstrate that the designed circuit exhibits antibunching, where the quadrature operator shows squeezing behavior.

**Introduction**

The squeezing state and its applications have been developed in recent decades [1-3]. It has been shown that squeezing originates from nonlinearity effects in any systems [1-7]. Different approaches and systems have been employed to generate squeezed state [3-5]. For instance, phase conjugate mirrors using four-wave mixing interactions have been applied to create a squeezed state [1,8]. Another important option is the use of a parametric amplifier, in which three-wave mixing is used to generate the squeezed state [5,7]. In addition, by controlling the spontaneous emission, a two-photon laser is applied to produce a squeezed state [1]. Moreover, atomic interaction with an optical wave can produce a nonlinear medium, leading to the creation of a squeezed state. Additionally, other phenomena such as third-order nonlinearity of the wave propagation in the optical fiber have the ability to generate a squeezed state [7]. In quantum applications, the squeezed state is very important because it introduces less fluctuation in one quadrature phase than the coherent state, which is very similar to the classical state [9-13]. Quantum fluctuation in a coherent state is equal to zero-point fluctuation, in which the standard quantum limits the reduction of the noise in a signal [1, 14]. Therefore, the noise fluctuation can be reduced below the standard limit when the system is in squeezed state [14]. There is no classical analog for the squeezed state, and this state, in contrast to the coherent state that shows Poisson photon counting (photon bunching) statistics, may show sub-Poisson photon counting (photon antibunching) [1,2, 13-14]. In other words, there is no direct connection between squeezing and photon antibunching, but each is a nonclassical phenomenon [1,3, 13]. For some quantum applications, such as quantum radar and quantum sensors [15-21] the noise effect is critical when the system tries to detect a very low-level signal. For signal detection, the received signals, which have very small levels, can be easily affected by noise. Therefore, to control and limit the noise, it is necessary to prepare the key subsystems, such as the low-noise amplifier (LNA) and detector, to operate in the squeezed state by which the noise can be reduced below the zero-point fluctuation. LNA is an electronic amplifier generally designed to amplify low-level signals while simultaneously keeping the noise at a very low level [23-25]. Today, the cryogenic LNA has been designed to operate at very low temperature to strongly limit the noise and so, due to this fact the cryogenic LNA is so popular in quantum applications [25-30].

With the knowledge of the points mentioned above, in this work, we attempt to design a circuit containing two external oscillators coupled to an InP high electron mobility transistor (HEMT) transistor operating at cryogenic temperature to create the squeezed states. This type of transistor was selected because HEMT technology does not have a strong effect from freeze-out at cryogenic temperature [25-28]. The designed circuit can be considered a core circuit for a low-noise amplifier used in front-end transceivers to amplify faint signals. In this study, the nonlinear properties of the cryogenic InP HEMT transistor play a key role, and we discuss how the emerging nonlinearity can affect the state of the coupling oscillators. Additionally, a critical point will be addressed which relates to the trade-off between squeezing generated by the nonlinear properties of the transistor and the degradation of the produced state because of the damping created by the transistor internal circuits.

**Theoretical and Backgrounds**

*System description*

The circuit is schematically shown in Fig. 1, which shows two LC oscillators (resonators) coupled to each other through a nonlinear device (depicted in the inset figure). As mentioned in the previous section, the main goal is to create a squeezing state in a low-noise amplifier (LNA), which is essential in quantum sensing applications [15-16, 26]. In fact, if such a circuit is prepared in the squeezing state, it helps minimize the noise effect. This implies that the performance of the cryogenic LNA, at which the noise strongly limits the operation, is enhanced. Therefore, the circuit shown in Fig.1 is designed. In this circuit, the transistor is used as a nonlinear element, and the state of the oscillators may exhibit squeezing. It has been theoretically shown that nonlinearity arises because the transistor can be expressed as a nonlinear capacitor, inductor coupling to the second oscillator, and second-order transconductance ($g_{m2-N}$). These factors are defined in detail in the next section, and can strongly manipulate the state of the oscillators to create squeezing. Additionally, Fig. 1 schematically reveals that only the second oscillator can generate squeezing in state; this important point will be discussed in detail later. Nonetheless, we theoretically demonstrate that coupling oscillators can generate two-mode squeezing.

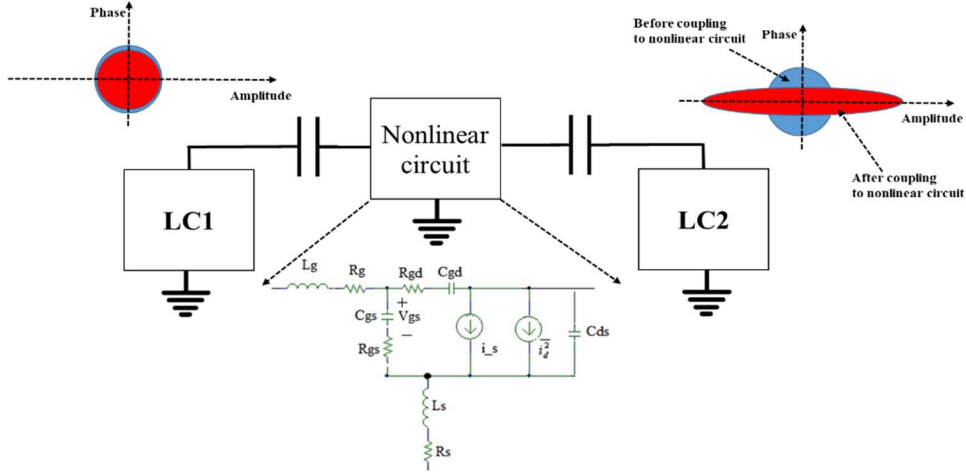

Fig. 1 Schematic of the system containing two external oscillators coupling through a nonlinear device, inset figure: nonlinear device internal circuit

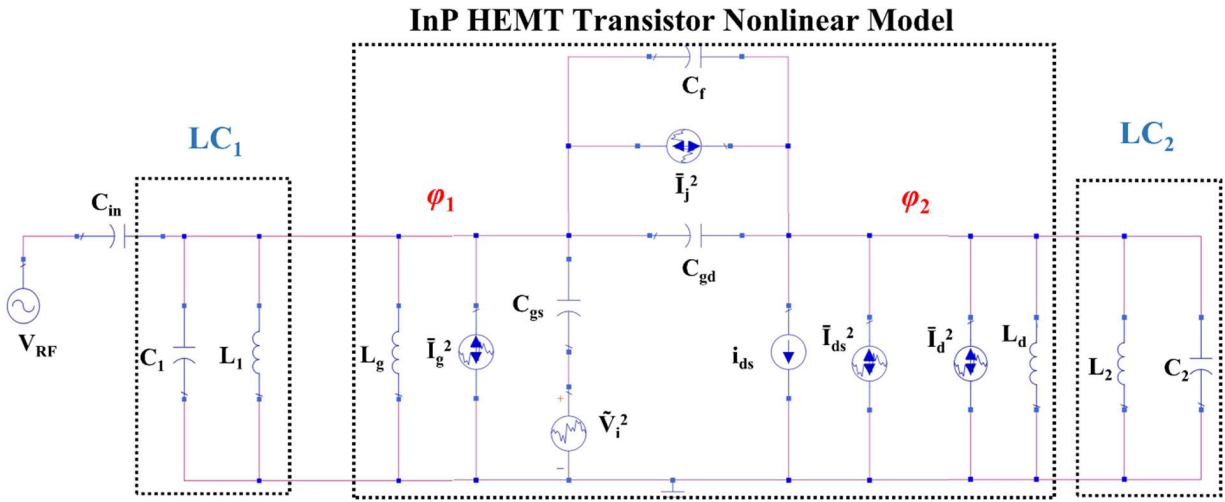

Fig. 2 Complete model of the system; $L_{C1}$ coupling to $L_{C2}$ through InP HEMT Transistor operating at 5K.

The high-frequency model of an InP HEMT transistor [26, 28] coupled with two oscillators is shown in Fig. 2. Some elements in the transistor nonlinear model are created owing to the high-frequency effect, such as $C_{gs}$, $C_{gd}$, $L_g$, and $L_d$ and some other elements, such as resistors, as shown in Fig. 1, are created to address the thermal loss in the circuit. These elements are sources of thermal noise in the transistor, and their effects are shown as voltage and current noise sources in Fig. 2. In fact, $\bar{I}_g^2$, $\bar{I}_j^2$, $\bar{I}_d^2$, and $\tilde{V}_i^2$ models the thermal noise of $R_g$, $R_{gd}$, $R_{ds}$, and $R_{gs}$, respectively. A noise model is applied to the circuit to demonstrate the effects of the contributed resistors. Additionally, we ignored the $L_s$ effect and merged $R_s$ with $R_{gs}$ to simplify algebra. In the circuit, $i_{ds}$ and $\bar{I}_{ds}^2$ are the dependent current sources containing the nonlinearity of the transistor and related thermal noise, which is a critical factor in generating noise. Finally, $C_f$, $C_{in}$, $V_{RF}$, $\varphi_1$ and $\varphi_2$ are the feedback capacitor and coupling capacitor used to separate the input signal from the DC signals, input signal, and node flux for the input and output nodes, respectively. This circuit is completely analyzed using quantum theory, and we will attempt to initially derive the contributed Lagrangian; then, the total Hamiltonian of the system is examined, in which the factors that cause the squeezing in state will be addressed.

*The designed system Hamiltonian*

In this section, the circuit shown in Fig.2 is analyzed using the full quantum theory. As shown in the circuit, it includes all noises that can affect the signals, such as the thermal noises generated by the dependent current source and resistors in the circuit. The data for the nonlinear model of the InP HEMT transistor operating at cryogenic temperatures are listed in Table. 1. First, we theoretically derive the total Lagrangian of the system to obtain the quantum properties of the circuit illustrated in Fig. 2. For the analysis, the coordinate variables are defined as $\varphi_1$ and $\varphi_2$ (node flux), as shown in Fig. 2, and the momentum conjugate variables are defined by $Q_1$ and $Q_2$ (loop charge). The total Lagrangian of the circuit is derived as [30]:

$$L_c = \frac{C_{in}}{2}(\dot{\varphi}_1 - V_{RF})^2 + \frac{C_1}{2}\dot{\varphi}_1^2 - \frac{1}{2L_1}\varphi_1^2 + \overline{I_g^2}\varphi_1 + \frac{C_{gs}}{2}(\dot{\varphi}_1 - \overline{V_i^2})^2 + \frac{C_{gd} + C_f}{2}(\dot{\varphi}_1 - \dot{\varphi}_2)^2 + i_{ds}\varphi_1 \\ (\overline{I_{ds}^2} + \overline{I_d^2})\varphi_2 + \frac{C_2}{2}\dot{\varphi}_2^2 - \frac{1}{2L_2}\varphi_2^2 + \overline{I_j^2}(\varphi_2 - \varphi_1) \tag{1}$$

In this equation the dependent current source is defined as $i_{ds} = g_m V_{gs} + g_{m2} V_{gs}^2 + g_{m3} V_{gs}^3$ [23,31], where $g_m$ is the intrinsic transconductance of the transistor and $g_{m2}$, $g_{m3}$ are the second- and third-order transconductance. These terms ($g_{m2}$, $g_{m3}$) bring nonlinearity in the circuit. Moreover, thermally generated noises by the resistors and the current source are defined as: $\bar{I}_g^2 = 4KT/R_g$, $\bar{I}_d^2 = 4KT/R_d$, $\bar{I}_j^2 = 4KT/R_j$, $\bar{I}_{ds}^2 = 4KT\gamma g_m$, and $\bar{I}_i^2 = 4KT/R_i$, where K, T, and γ respectively are the Boltzmann's constant and operational temperature, and empirical constant [30]. Using the Legendre transformation [14, 30], one can theoretically derive the classical Hamiltonian of the circuit. For this, it is necessary to calculate the momentum conjugate variables using $Q_i = \partial L_c / \partial(\partial \varphi_i / \partial t)$ for i = 1,2 represented as:

$$Q_1 = (C_{in} + C_1 + C_{gs} + C_f + C_{gd})\dot{\varphi}_1 - (C_f + C_{gd})\dot{\varphi}_2 - C_{in}V_{RF} + g_m\varphi_2 + 2g_{m2}\varphi_2\dot{\varphi}_1 + 3g_{m3}\varphi_2\dot{\varphi}_1^2 - C_{gs}\overline{V_i^2} \\ Q_2 = (C_2 + C_f + C_{gd})\dot{\varphi}_2 - (C_f + C_{gd})\dot{\varphi}_1 \tag{2}$$

Applying Legendre transformation, the classical Hamiltonian is expressed as:

$$H_c = \left\{ \frac{C_A}{2}\dot{\varphi}_1^2 + \frac{1}{2L_1}\varphi_1^2 + \frac{C_B}{2}\dot{\varphi}_2^2 + \frac{1}{2L_2}\varphi_2^2 \right\} + \left\{ -C_c\dot{\varphi}_1\dot{\varphi}_2 + g_{m2}\varphi_2\dot{\varphi}_1^2 + 2g_{m3}\varphi_2\dot{\varphi}_1^3 \right\}$$
$$+ \left\{ -\varphi_1\left(\overline{I_g^2} - \overline{I_j^2}\right) - \varphi_2\left(\overline{I_d^2} + \overline{I_j^2} + \overline{I_{ds}^2}\right) - \frac{C_{gs}}{2}\overline{V_i^2} \right\} - \frac{C_{in}}{2}V_{RF}^2 \tag{3}$$

where $C_A = C_{in}+C_1+C_{gs}+C_f+C_{gd}$, $C_B = C_{gd} + C_f + C_2$, and $C_C = C_f + C_{gd}$. In Eq. 3, the first term relates to the LC resonance Hamiltonian affected by the coupling elements. It is clearly shown that $C_A$ and $C_B$ are affected due to the transistor internal circuit. The second term contributes to the linear and nonlinear coupling between the LC resonators and the nonlinear circuit, and finally the third term defines the noise effect in the system which is originally generated by the transistor nonlinearity in the circuit. In the following, using Eq. 2, one can express the first derivative of the coordinate variables ($\partial\varphi_i/\partial t$) in terms of the momentum conjugates ($Q_i$) represented in the matrix form as:

$$\begin{bmatrix} \dot{\varphi}_1 \\ \dot{\varphi}_2 \end{bmatrix} = \frac{1}{C_M^2} \left\{ \begin{bmatrix} C_B & C_C \\ C_C & C_A+C_N \end{bmatrix} \begin{bmatrix} Q_1 \\ Q_2 \end{bmatrix} + \begin{bmatrix} C_B & C_C \\ C_C & C_A+C_N \end{bmatrix} \begin{bmatrix} C_{in} & 0 \\ 0 & 0 \end{bmatrix} \begin{bmatrix} V_{RF} \\ 0 \end{bmatrix} - \begin{bmatrix} C_B & C_C \\ C_C & C_A+C_N \end{bmatrix} \begin{bmatrix} 0 & g_m \\ 0 & 0 \end{bmatrix} \begin{bmatrix} \varphi_1 \\ \varphi_2 \end{bmatrix} \right\} \tag{4}$$

where $C_M^2 = C_B(C_A+C_N)-C_C^2$ and $C_N = 2g_{m2}\varphi_{2\_dc} + 6g_{m3}\varphi_{2\_dc}(\partial\varphi_1/\partial t)|_{dc}$ is the capacitor generated due to the nonlinearity effect. Following, we will show that this quantity strongly affects the coupled LC resonators frequency and impedances and consequently the quantum properties of the LC resonators is severely influenced by $C_N$. By substitution of Eq. 4 into Eq. 3 one can derive the total Hamiltonian for the system; however, to study in detail about the system and getting to know about each factor's impact in the system, we initially use the linearization technique to linearize the nonlinear terms in the second term of Eq.3 and derive the linear Hamiltonian of the system as: Thus, the linear Hamiltonian of the system is defined as:

$$H_L = \left\{ \frac{1}{2C_{q1}}Q_1^2 + \frac{1}{2L_1}\varphi_1^2 + \frac{1}{2C_{q2}}Q_2^2 + \frac{1}{2L_2'}\varphi_2^2 \right\}$$
$$\left\{ \frac{1}{2C_{q1q2}}Q_1Q_2 + g_{12}Q_1\varphi_2 + g_{22}Q_2\varphi_2 \right\} + \left\{ V_{q1}Q_1 + V_{q2}Q_2 + I_{p2}\varphi_2 - \overline{I_{gs}^2}\varphi_1 \right\} \tag{5}$$

where $C_{q1}$, $C_{q2}$, $C_{q1q2}$, $g_{12}$, $g_{22}$, $L_2'$, $V_{q1}$, $V_{q2}$, and $I_{p2}$ are defined in Appendix A and the dc terms are ignored for simplicity. In fact, these coefficients are essentially the function of $g_m$, $C_N$, $C_C$, and $V_{RF}$ by which the nonlinearity effect of the transistor is emphasized. In other words, the nonlinearity created by the transistor induces some factors by which the properties of the coupling LC resonators are strongly affected. For instance, the LC resonators impedances are $Z_1 = \sqrt{L_1/C_{q1}}$, $Z_2 = \sqrt{L_2'/C_{q2}}$, and the associated frequencies are $\omega_1 = 1/\sqrt{L_1 C_{q1}}$, $\omega_2 = 1/\sqrt{L_2' C_{q2}}$. The relations show that the coupling oscillator's impedance and frequencies specially the second LC becomes affected. Additionally, some terms in Eq.2 such as $Q_1Q_2$, $Q_2\varphi_1$, and $Q_1\varphi_2$ show the coupling between oscillators in the circuit design. Also, in the third term in Eq.5 some terms such as $V_{q1}Q_1$, $\overline{I_{gs}^2}\varphi_1$, $V_{q2}Q_2$, $I_{p2}\varphi_2$ in the equation declare the RF source and thermal noise coupling to the contributed oscillators. In this equation, $\overline{I_{gs}^2} = \overline{I_g^2} - \overline{I_j^2}$. In the following, one can derive the linear Hamiltonian in terms of annihilation and creation operators using the quantization procedure for the coordinates and the related momentum conjugates. The quadrature operators defined as $Q_1 = -i(a_1-a_1^+)\sqrt{\hbar/2Z_1}$, $\varphi_1 = (a_1+a_1^+)\sqrt{\hbar Z_1/2}$ and $Q_2 = -i(a_2-a_2^+)\sqrt{\hbar/2Z_2}$, $\varphi_2 = (a_2+a_2^+)\sqrt{\hbar Z_2/2}$, where $(a_i,a_i^+)$ $i = 1,2$ are the first and second oscillator's annihilation and creation operators. Thus, the linear Hamiltonian in terms of the ladder operators is given by:

$$H_L = \left\{\hbar\omega_1\left(a_1^+a_1+\frac{1}{2}\right)+\hbar\omega_2\left(a_2^+a_2+\frac{1}{2}\right)\right\}$$
$$+\left\{-\frac{\hbar}{2}\frac{1}{C_{q1q2}\sqrt{Z_1Z_2}}(a_1-a_1^+)(a_2-a_2^+)-\frac{i\hbar}{2}g_{12}\sqrt{\frac{Z_2}{Z_1}}(a_1-a_1^+)(a_2+a_2^+)-\frac{i\hbar}{2}g_{22}(a_2-a_2^+)(a_2+a_2^+)\right\} \quad (6)$$
$$+\left\{-iV_{q1}\sqrt{\frac{\hbar}{2Z_1}}(a_1-a_1^+)-iV_{q2}\sqrt{\frac{\hbar}{2Z_2}}(a_2-a_2^+)+I_{p2}\sqrt{\frac{\hbar Z_2}{2}}(a_2+a_2^+)-\overline{I_{gs}^2}\sqrt{\frac{\hbar}{2Z_2}}(a_1+a_1^+)\right\}$$

In the following, it is necessary to add the nonlinearity to the Hamiltonian and derive the total Hamiltonian containing the linear and nonlinear parts. The nonlinear terms in Eq. 3 can be re-written as:

$$H_N = \{g_{m2}+6g_{m3}\dot\varphi_{1\_dc}\}\varphi_2\dot\varphi_1^2 \quad (7)$$

Using Eq. 4, the nonlinear Hamiltonian is given by:

$$H_N = g_{m2\_N}\left[\left\{\frac{C_B^2}{C_M^4}\varphi_2Q_1^2+\frac{C_C^2}{C_M^4}\varphi_2Q_2^2+\frac{2C_BC_C}{C_M^4}\varphi_2Q_2Q_1-\frac{2g_mC_B^2}{C_M^4}\varphi_2^2Q_1-\frac{2g_mC_BC_C}{C_M^4}\varphi_2^2Q_2+\frac{g_m^2C_B^2}{C_M^4}\varphi_2^3\right\}_{NL}\right.$$
$$\left.\left\{\underbrace{\frac{-2g_mC_B^2C_{in}V_{RF}}{C_M^4}}_{1/2L_{2N}}\varphi_2^2+\underbrace{\frac{2C_B^2C_{in}V_{RF}}{C_M^4}}_{g_{12N}}Q_1\varphi_2+\underbrace{\frac{2C_BC_CC_{in}V_{RF}}{C_M^4}}_{g_{22N}}Q_2\varphi_2+\underbrace{\frac{C_B^2V_{in}^2V_{RF}^2}{C_M^4}}_{I_{p2N}}\varphi_2\right\}_L\right] \quad (8)$$

where $g_{m2\_N} = g_{m2} + 6g_{m3}(\partial\varphi_1/\partial t)|_{dc}$. The linear part of the Eq. 8 can directly attach to Eq. 5 to make the modified linear Hamiltonian given by:

$$H_L = \left\{\frac{1}{2C_{q1}}Q_1^2+\frac{1}{2L_1}\varphi_1^2+\frac{1}{2C_{q2}}Q_2^2+\left(\frac{1}{2L_2}+\frac{1}{2L_{2N}}\right)\varphi_2^2\right\}$$
$$\left\{\frac{1}{2C_{q1q2}}Q_1Q_2+(g_{12}+g_{12N})Q_1\varphi_2+(g_{22}+g_{22N})Q_2\varphi_2\right\}+\left\{V_{q1}Q_1+V_{q2}Q_2+(I_{p2}+I_{p2N})\varphi_2-\overline{I_{gs}^2}\varphi_1\right\} \quad (9)$$

As clearly seen in Eq. 9, the nonlinear Hamiltonian can change the coupling between oscillators in the circuit and the most important factor is $g_{m2\_N}$ by which the coupling between different coordinates and their momentum conjugates are manipulated. Additionally, attachment from Eq. 8 to Eq. 5 leads to change the second resonator's inductance by factor of $L_{2N}$. The term brought from nonlinearity ($L_{2N}$) manipulates the second resonator's impedance and frequency. Finally, the nonlinear terms of Eq. 8 are considered and so, the total Hamiltonian of the system in the terms of the ladder operators is given by:

$$H_t = \left[ \left\{ \hbar\omega_1 \left( a_1^+ a_1 + \frac{1}{2} \right) + \hbar\omega_2 \left( a_2^+ a_2 + \frac{1}{2} \right) \right\} \right.$$

$$+ \left\{ -\frac{\hbar}{2} \frac{1}{C_{q1q2}\sqrt{Z_1 Z_2}} (a_1 - a_1^+)(a_2 - a_2^+) - \frac{i\hbar}{2} g_{12}' \sqrt{\frac{Z_2}{Z_1}} (a_1 - a_1^+)(a_2 + a_2^+) - \frac{i\hbar}{2} g_{22}' (a_2 - a_2^+)(a_2 + a_2^+) \right\}$$

$$\left. + \left\{ -iV_{q1} \sqrt{\frac{\hbar}{2Z_1}} (a_1 - a_1^+) - iV_{q2} \sqrt{\frac{\hbar}{2Z_2}} (a_2 - a_2^+) + I_{p2}' \sqrt{\frac{\hbar Z_2}{2}} (a_2 + a_2^+) - \overline{I_{gs}^2} \sqrt{\frac{\hbar}{2Z_2}} (a_1 + a_1^+) \right\} \right]_L \quad (10)$$

$$+ \left[ -\hbar g_{13} (a_1 - a_1^+)^2 (a_2 + a_2^+) + \hbar g_{14} (a_2 + a_2^+)(a_2 - a_2^+)^2 - \hbar g_{15} (a_1 - a_1^+)(a_2 - a_2^+)(a_2 + a_2^+) \right.$$

$$\left. + \hbar g_{16} (a_2 + a_2^+)^3 + i\hbar g_{17} (a_1 - a_1^+)(a_2 + a_2^+)^2 + i\hbar g_{18} (a_2 - a_2^+)(a_2 + a_2^+)^2 \right]_{NL}$$

where $I_{P2'} = I_{P2} + I_{P2N}$, $g_{12}' = g_{12} + g_{12N}$, and $g_{22}' = g_{22} + g_{22N}$. Also, $g_{13} = (1/2Z_1)(\sqrt{(\hbar/2Z_2)})g_{m2\_N}C_B^2/C_M^4$, $g_{14} = (1/2Z_2)(\sqrt{(\hbar/2Z_2)})g_{m2\_N}C_C^2/C_M^4$, $g_{15} = (1/\sqrt{Z_2 Z_1})(\sqrt{(\hbar/2Z_2)})g_{m2\_N}C_B C_C/C_M^4$, $g_{16} = (Z_2/2)(\sqrt{(\hbar Z_2/2)})g_m^2 g_{m2\_N} C_B^2/C_M^4$, $g_{17} = Z_2(\sqrt{(\hbar/2Z_1)})g_m g_{m2\_N} C_B^2/C_M^4$, and $g_{18} = Z_2(\sqrt{(\hbar/2Z_2)})g_m g_{m2\_N} C_B C_C/C_M^4$. It is clearly shown in coefficients from $g_{13}$ to $g_{18}$ in which the effect of $g_{m2\_N}$ is dominant. In other words, the nonlinearity of the system in this work is strongly changed and controlled by the second-order transconductance $g_{m2\_N}$. Now, one can show using the total Hamiltonian of the system in which what terms in the presented Hamiltonian in Eq. 10 has the ability to generate the squeezing state.

*Generation of the squeezing state*

A squeezed-coherent state is generally produced by acting of the squeezed and displacement operators in, on the vacuum state defined mathematically as $|\alpha, \zeta\rangle = D(\alpha)S(\zeta) |0\rangle$, where $|0\rangle$ is the vacuum state [14]. It is found that the coherent state is generated by the linear terms in the Hamiltonian, whereas a squeezed state needs quadratic terms such as $a^2$ and $a^{+2}$ in the exponent. The squeezed-coherent state $S(\zeta,\alpha)$ can be analyzed as the evolution $\exp[Ht/j\hbar]$ under the Hamiltonian defined in Eq. 10. Based on this definition, any quadratic terms such as $a^2$ and $a^{+2}$ in the Hamiltonian may generate squeezing. The Hamiltonian in Eq. 10, the squeezed state can be generated by:

$$S(\zeta) = \exp \left[ \zeta_1 (a_2^2 - a_2^{+2}) + \left( \frac{\zeta_2^*}{2} a_2^2 - \frac{\zeta_2}{2} a_2^{+2} \right) \right] t \quad (11)$$

In this equation the squeezing parameters are defined as $\zeta_1 = -0.5 g_{22}' + g_{18} Re\{A_2\} + j g_{14} Im\{A_2\}$ and $\zeta_2 = 2A_1^*(-g_{17} - j g_{15})$, where $A_1$ and $A_2$ are the strong fields (DC points) of $LC_1$ and $LC_2$. The DC points can be calculated using Heisenberg-Langevin equations in the steady-state [15, 16]. Also $Re\{\}$ and $Im\{\}$ indicate the real and imaginary parts, respectively. Eq. 11 clearly shows that the squeezing is generated just for $LC_2$ and it does not happen for $LC_1$. This point contributes to the nonlinearity terms in Hamiltonian expressed in Eq. 10 and also is related to the dependent current source containing $g_{m2}$ and $g_{m3}$, which is directly connected to $LC_2$. The important point about the squeezing strength parameters $\zeta_1$ and $\zeta_2$ is that each of them is dependent on $g_{m2\_N}$. In Eq. 11, $\zeta_1$ and $\zeta_2$ are complex numbers which means that the squeezing parameters contain the phase which determines the angle of quadrature to be squeezed. Additionally, we found that the system can generate two-mode squeezing. That means that the nonlinearity created by the transistor couples two oscillators in such a way that the coupled modes become squeezed. The expression generated due to the Hamiltonian of the system for two-mode squeezing is expressed as:

$$S_2(\zeta) = \exp\left[\left(\frac{\zeta_{t1}^*}{2}a_1a_2 - \frac{\zeta_{t1}}{2}a_1^+a_2^+\right) + \left(\frac{\zeta_{t2}^*}{2}a_1a_2 - \frac{\zeta_{t2}}{2}a_1^+a_2^+\right)\right]t \tag{12}$$

where $\zeta_{t1} = jA_2g_{15}$ and $\zeta_{t2} = jA_1g_{13}$. In the same way, the squeezing parameters are strongly dependent on $g_{m2\_N}$. Finally, one can easily find that the system can generate the coherent state which means that the state generated by the Hamiltonian expressed in Eq. 10 is a squeezed-coherent state or a two-mode squeezed-coherent state. In the following, we just focus on the squeezed-coherent state and study the parameters that can manipulate the squeezing states. For simulation, "t" in the exponent ($\exp[Ht/j\hbar]$) is defined as $t_0 < \{1/\kappa_1, 1/\kappa_2\}$, where $\kappa_1$ and $\kappa_2$ are the first and second oscillators decay rate. By selecting $t_0$, the system is forced to generate squeezing before the resonator decaying by which the squeezing is destroyed [1]. Some more information is introduced about $t_0$ in Appendix B.

Table. 1 Values for the small signal model of the 2*50 μm InP HEMT at 5 K [32-34].

|  | **Stands for** | **Value Unit** |
|---|---|---|
| $R_g$ | Gate resistance | 0.3 Ω |
| $L_g$ | Gate inductance | 75 pH |
| $L_d$ | Drain inductance | 70 pH |
| $C_{gs}$ | Gate-Source capacitance | 69 fF |
| $C_{ds}$ | Drain-Source capacitance | 29 fF |
| $C_{gd}$ | Gate-Drain capacitance | 19 fF |
| $R_{gs}$ | Gate-Source resistance | 4 Ω |
| $R_{gd}$ | Gate-Drain resistance | 35 Ω |

In this study, quadrature variance is used to demonstrate the behavior of the state generated by the oscillators. In addition, we used the bunching and antibunching behavior of the generated photons. The second-order correlation function, $g^2(\tau)$, must be calculated. For the designed system, with regard to the fact that the system is limited by t0, the photon counting time is sufficiently short. Thus, for such a short counting time, the variance of the photon number distribution is related to the second-order correlation function $g^2(\tau = 0) = 1 + (V(n) - <n>)/<n>$, where $V(n)$ and $<n>$ are the photon number variance and the average, respectively [1, 13]. It has been shown that for light with sub-Poissonian statistics $g^2(\tau = 0) < 1$, this phenomenon is called anti-bunching, which is a nonclassical phenomenon. Of course, $g^2(\tau = 0) < 1$ is not a necessary condition for squeezing the state; however, if $g^2(\tau = 0) > 1$, the field is a classical field [1]. In other words, the squeezing state may exhibit bunching and antibunching behaviors. In the following section, the aforementioned points are discussed with some related simulations.

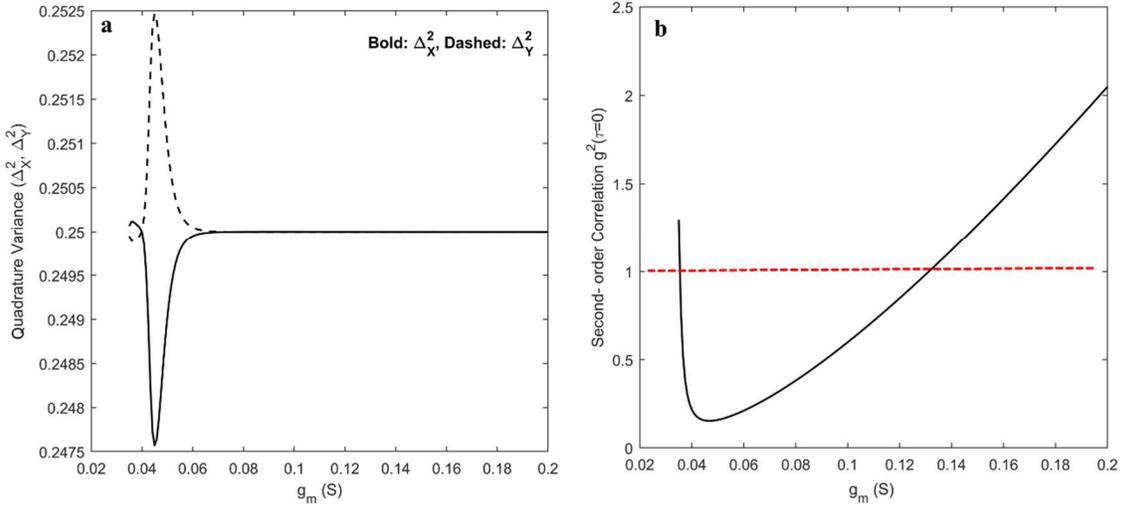

Fig. 3 a) Quadrature variance vs. $g_m$ (S), b) Second-order correlation function vs. $g_m$ (S); $C_f = 0.02$ pF, $g_{m3}$ = 1200 mA/V$^3$, $g_{m2}$ = 200 mA/V$^2$, $V_{RF}$ = 1 µV, $\kappa = \kappa_1 = \kappa_2 = 0.001\omega$.

The squeezing of the second oscillator with respect to Eq. 11 is simulated and the results are shown in Fig. 3. As seen in Fig. 3a, which illustrates the quadrature operator's variance versus $g_m$, the squeezing appears in the resonator and reaches to the maximum value for $g_m$ around 45 mS. However, the amount of the squeezing is decreased when $g_m$ exceeds 60 mS. It may relate to the fact that $g_m$ directly manipulates $\bar{I}_{ds}^2$ by which the noise exceeds in the system and due to that, the squeezing is strongly limited when $g_m$ is increased. Additionally, in Fig. 3a, the dashed graph shows $\Delta y^2 > 0.25$, indicating that this operator for each value of $g_m$ shows bunching; in other words, the operator shows classical field behavior. We theoretically show that the important factors affecting LC$_2$ to generate squeezing state include $C_N$, $C_N'$, and $g_{m2\_N}$. To get know about the factors effects and make a comparison with other elements, the contributed values are calculated for $g_m$ = 45 mS and represented as: $C_N$ = 3.3pF, $C_N'$ = 72 mA/V, and $g_{m2\_N}$ = 677 mA/V$^2$. These values show nonlinear effects in the transistor which changes the electrical properties of the circuit. For instance, one can compare $C_N$ with $C_{gd}$ or $C_{gs}$, which indicates that $C_N$ is greater than the internal capacitances and $g_{m2\_N}$ is comparable with $g_{m2}$.

In addition, the second-order correlation function $g^2(\tau = 0)$ behavior can be considered, as illustrated in Fig. 3b. The figure shows perfect consistency with the quadrature operators' variance around $g_m$ = 40 mS. The value of $g^2(\tau = 0)$ around 45 mS reaches its minimum and is less than 1 which means that the second resonator exhibits antibunching. Notably, the change in the sign of $g^2(\tau = 0)$ from bunching ($g^2(\tau = 0) > 1$) to antibunching ($g^2(\tau = 0) < 1$) indicates squeezing in the system. The results shown in Fig. 3b reveal that squeezing occurs only for small values of $g_m$. In other words, the current amplification factor ($g_m$) in the transistor should be maintained at a low level to generate squeezing at cryogenic temperatures. Nonetheless, this is clearly shown in Eq. 11 that $\zeta_1$ and $\zeta_2$ are strongly manipulated by $g_{m2-N}$ which is a fundamental function of $g_{m2}$, $g_{m3}$, and that $C_N$ plays a key role in changing the coupling between resonators. Additionally, other factors such as feedback capacitance, LC resonator decay rate, and input RF source can influence the squeezing in the system. For instance, the effect of $g_{m3}$ as a nonlinearity factor on the quadrature operator variance and photon bunching and antibunching is illustrated in Fig. 4. Fig. 4a shows that by increasing $g_{m3}$ the quadrature variance increases. This contributed to the increase in the squeezing strength parameters. In addition, the figure shows that increasing $g_{m3}$ leads to maintaining $\Delta x^2 < 0.25$ for larger $g_m$. In the same way, the effect of $g_{m3}$ increasing on the second-order correlation function is depicted in Fig. 4b. This reveals that increasing $g_{m3}$ causes an increase in $g_m$ to 120 mS, in which the second-order correlation shows antibunching. This contributes to the fact that increasing $g_{m3}$ changes $C_N$ and $g_{m2-N}$ leading to a strengthening the squeezing behavior.

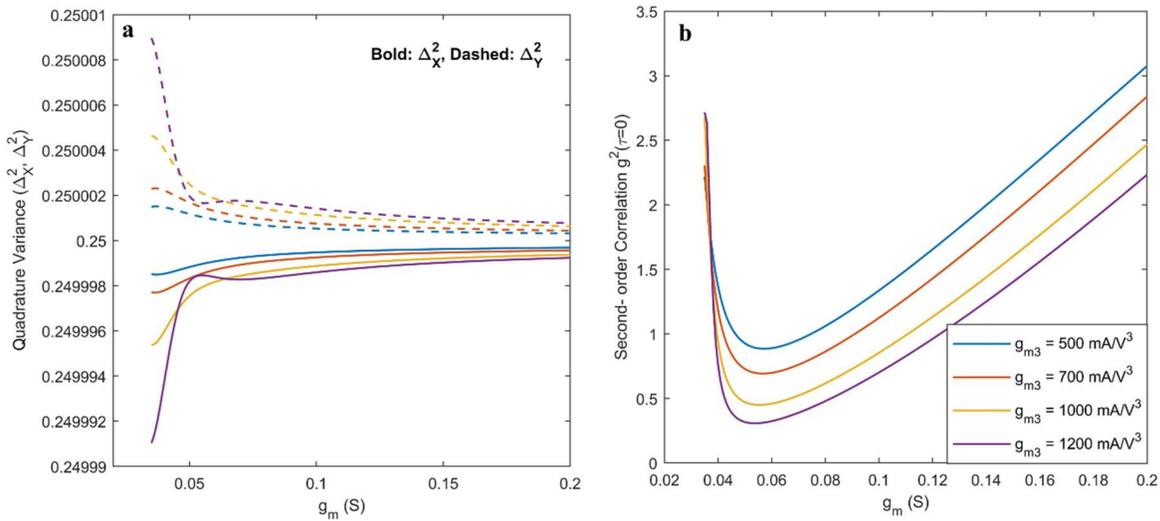

Fig.4 Quadrature variance vs. $g_m$ (S), b) Second-order correlation function vs. $g_m$ (S) for different $g_{m3}$(mA/V$^3$); $C_f$ = 0.02 pF, $g_{m2}$ = 200 mA/V$^2$, $V_{RF}$ = 1 μV, $\kappa$ = 0.001ω.

Additionally, in this study, the effects of other parameters such as $C_f$, $V_{RF}$, $g_{m2}$, and oscillator decay rate $\kappa$ are analyzed, and the results of the simulations are depicted in Fig. 5. In this graph, the red dashed line is inserted to easily trace the bunching to antibunching (and vice versa) entry point as a function of $g_m$. In this simulation, it is assumed that the two oscillators had the same decay rate $\kappa_1 = \kappa_2 = \kappa$. As expected, Fig. 5 shows that increasing $C_f$, $V_{RF}$, and $g_{m2}$ causes an increase in antibunching, whereas an increase in the decay rate leads to a decrease in antibunching. In this figure, the key factor that can be freely manipulated it, is the feedback capacitor, by which circuit properties, such as noise, gain, and stability, can be manipulated. The graph in Fig. 5b reveals that increasing the feedback capacitor causes antibunching for a larger gm. This may be related to the noise figure enhancement using feedback in the circuit. In other words, using a feedback capacitor strongly enhances the noise figure of the circuit, which means that eliminating noise leads to enhanced squeezing.

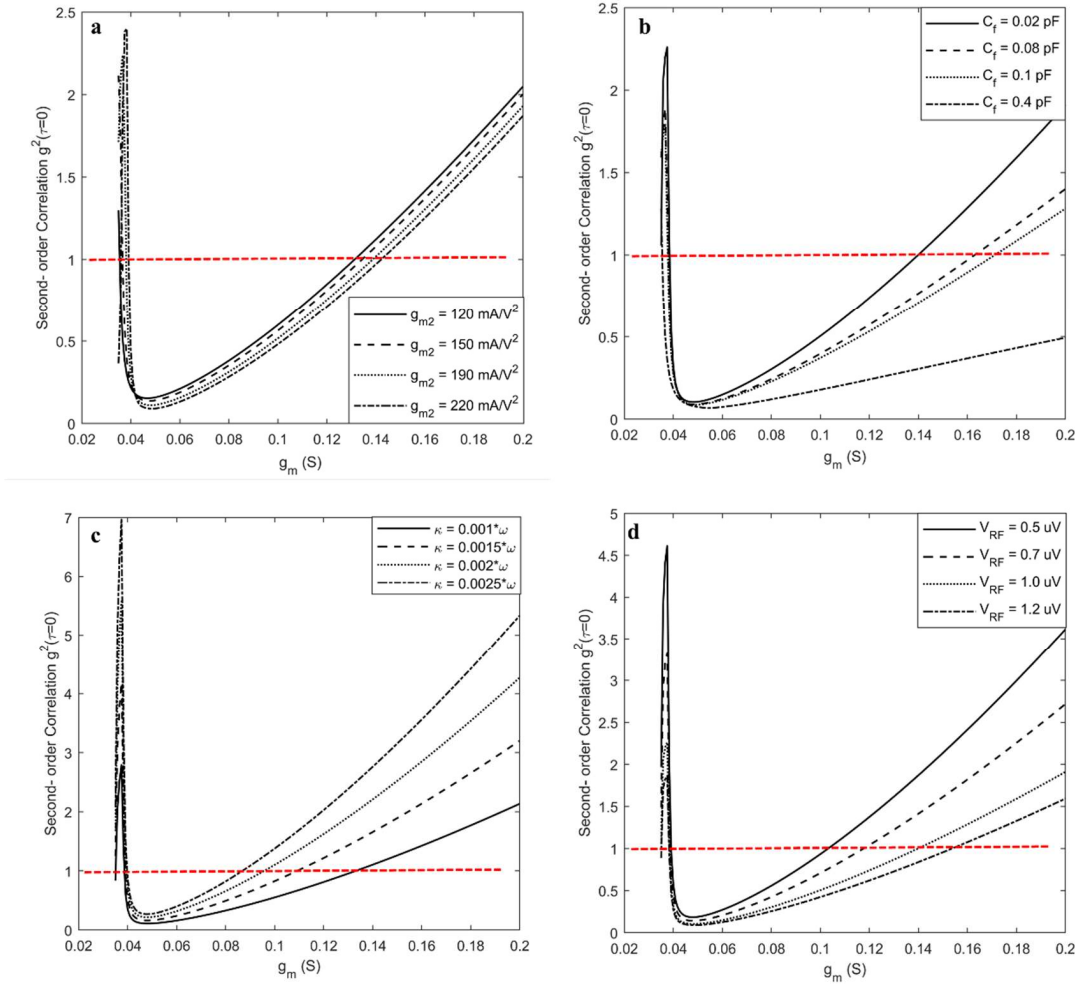

Fig.4 Second-order correlation function vs. $g_m$ (S) a) $g_{m2}$ effect, b) feedback capacitance effect ($C_f$), C) LC resonator decay rate effect ($\kappa$), d) Input RF source amplitude effect ($V_{RF}$); $g_{m3} = 1200$ mA/V$^3$.

The results illustrated in this study show that cryogenic InP HEMT transistor nonlinearity has the ability to generate a squeezed state. This is an important achievement, because such a system can be an essential part of a cryogenic detector that has been used in quantum applications. Thus, the operation of the detector or amplifier in the squeezed state implies that the noise fluctuation is limited below the zero-point fluctuations. This is an interesting goal of this study; nonetheless, we know that this is challenging to achieve.

**Conclusions**

This article mainly emphasizes the generation of the squeezing state using the nonlinearity of the InP HEMT transistor. For this purpose, we designed a circuit containing two external oscillators coupled with a cryogenic InP HEMT transistor operating at 5 K. The circuit was analyzed using quantum theory, and the contributions of the Lagrangian and Hamiltonian functions were theoretically derived. Some key factors in the Hamiltonian arise because the nonlinearity of the transistor can generate a squeezed state. Thus, we focused on these parameters and their engineering attempts to generate squeezing. The results show that the squeezed state occurred only for the second oscillator. This implies that the first oscillator experiences a coherent state. In addition, we theoretically demonstrate that two coupled oscillators through a cryogenic transistor have the ability to generate two-mode squeezing. Thus, as a general point, if such a cryogenic circuit has the ability to generate squeezing, then the critical noise fluctuations would be strongly limited. This means that important phenomena, such as entanglement, can persist for a long time in the quantum transceiver.

## Appendixes:

**Appendix A:**

In this appendix all of the parameters used in the main article listed as $C_{q1}$ [F], $C_{q2}$ [F], $C_{q1q2}$ [F], $g_{12}$ [1/s], $g_{22}$ [1/s], $V_{q1}$ [v], $V_{q2}$ [v] and $I_{P2}$ [A] are given by:

$$\frac{1}{C_{q1}} = \frac{2C_B^2(C_N + 0.5C_A)}{C_M^4} - \frac{C_C^2 C_B}{C_M^4}$$

$$\frac{1}{C_{q2}} = \frac{2C_C^2(C_N + 0.5C_A)}{C_M^4} + \frac{C_A'^2 C_B}{C_M^4} - \frac{2C_C(C_N + C_A)C_B}{C_M^4}$$

$$\frac{1}{C_{q1q2}} = \frac{2C_C C_B(C_N + 0.5C_A)}{C_M^4} + \frac{C_C(C_N + C_A)C_B}{C_M^4} - \frac{C_C^2 C_B}{C_M^4}$$

$$\frac{1}{L_{p2}} = \frac{2g_m^2 C_B^2(C_N + 0.5C_A)}{C_M^4} + \frac{2g_m C_N' C_B}{C_M^2}$$

$$g_{12} = \frac{-2g_m C_B^2(C_N + 0.5C_A)}{C_M^4} + \frac{C_N' C_B}{C_M^2} - \frac{3g_m C_C^2 C_B}{2C_M^4}$$

$$g_{22} = \frac{-2g_m C_B C_C(C_N + 0.5C_A)}{C_M^4} + \frac{C_N' C_C}{C_M^2} + \frac{g_m C_C^3}{C_M^4} - \frac{g_m(C_N + C_A)C_C C_B}{C_M^4}$$

$$V_{q1} = \frac{2C_B C_C C_{in} V_{RF}(C_N + 0.5C_A)}{C_M^4} - \frac{C_C^2 C_{in} C_B V_{RF}}{C_M^4}$$

$$V_{q2} = \frac{2C_B C_C C_{in} V_{RF}(C_N + 0.5C_A)}{C_M^4} - \frac{C_C^3 C_{in} V_{RF}}{C_M^4}$$

$$I_{p2} = \frac{-2g_m C_B^2 C_{in} V_{RF}(C_N + 0.5C_A)}{C_M^4} + \frac{g_m C_B C_{in} C_C^2 V_{RF}}{C_M^4} - \frac{C_B C_{in} C_N' V_{RF}}{C_M^2} - \overline{I_{ds}^2}$$

where and $C_N' = 2g_{m2}(\partial\varphi_1/\partial t)|_{dc} + 12g_{m3}[(\partial\varphi_1/\partial t)|_{dc}]^2$.

**Appendix B:**

In this appendix, we try to give some information about $t_0$. In the main article, it is discussed that $t_0$ is selected less than the times that two oscillators' decays with it. In fact, from classical point of view, $t_0$ should be in the order of the steady-state time. Therefore, in this part we tried to calculate the step response of the circuit. For this reason, however for simplicity, a simplified version of the circuit shown in Fig. 2 is demonstrated in Fig. B1 and the related transfer function is derived as:

$$\frac{V_{out}(s)}{V_{RF}(s)} = \frac{g_m L_{p2} L_1 S^2}{L_{p2} L_1 C_{p1} C_{p2} S^4 + L_{p2} L_1 C_{p2} S^3 + (L_1 C_{p1} + L_{p2} C_{p2}) S^2 + L_1 S + 1} \tag{B1}$$

where $L_{p2} = L_{2N} \| L_2$, $C_{p1} = C_{gs} + C_1 + (C_{gd}+C_f).A_{v0}$, $C_{p2} = C_N + C_2$. As can be seen in the expressions, the second oscillator's inductance and capacitance are affected by the nonlinearity effects as $L_{2N}$ and $C_N$ and the first oscillator is just influenced by the gain of the circuit $A_{v0} \sim g_m r_0$, where $r_0$ is the resistance generated due to the channel length modulated effect. The step response of the transfer function expressed in Eq. B1 is illustrated in Fig. B2. It is shown that the settling time is around 80 nsec; this time is very close to $t_0$ that we selected based on the oscillators decay rates. In fact, $t_0$ is selected around the settling time for the system.

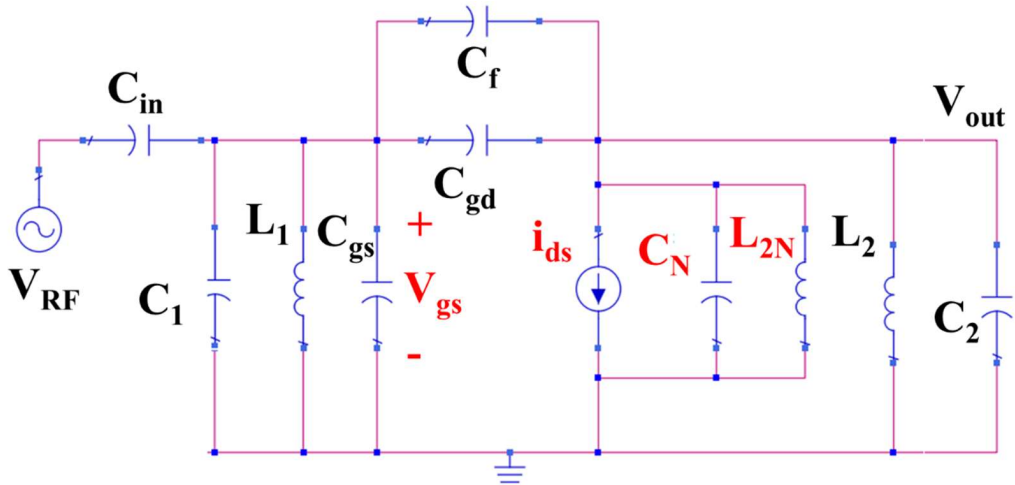

Fig. B1 A simplified version of the circuit to estimate about the transfer function.

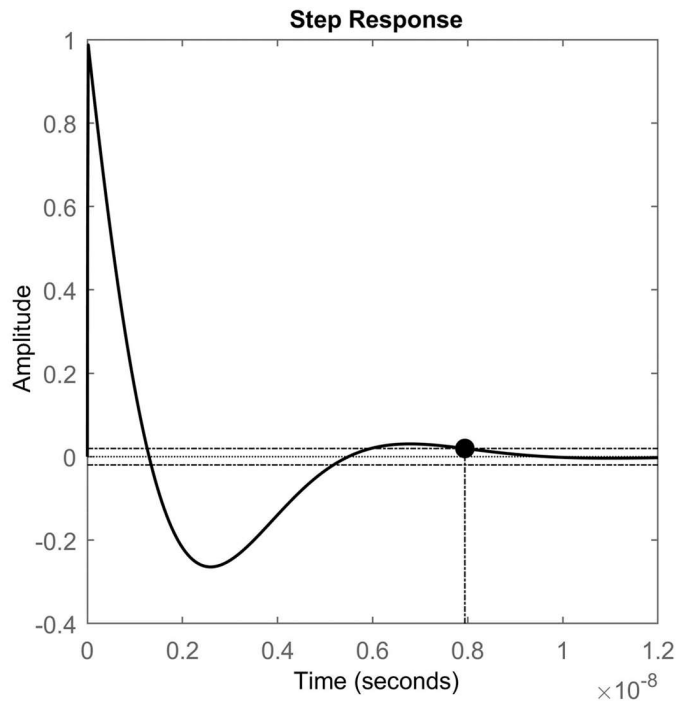

Fig. B2 Step response of the circuit.